\documentclass[usenatbib]{mn2e}
\usepackage{graphicx, amssymb, aas_macros, natbib}
\newcommand{\vmax}{\rm V_{max}}
\newcommand{\vinfall}{\rm V_{infall}}

\newcommand{\mstar}{\rm \, M_{\star}}
\newcommand{\msun}{\rm \, M_{\odot}}

\newcommand{\hmpc}{h^{-1} \, \rm Mpc}
\newcommand{\hkpc}{h^{-1} \, \rm kpc}
\newcommand{\mpc}{\rm \, Mpc}
\newcommand{\kpc}{\rm \, kpc}
\newcommand{\gyr}{\rm \, Gyr}

\newcommand{\kms}{{\rm km \, s}^{-1}}

\newcommand{\msii}{MS-II}

\title[The surprising inefficiency of dwarf satellite quenching]
{
The surprising inefficiency of dwarf satellite quenching
}
\author[C. Wheeler et al.]{Coral Wheeler$^1$\thanks{$\!$crwheele@uci.edu},
  John I. Phillips$^1$, Michael C. Cooper$^1$, Michael Boylan-Kolchin$^{2}$,
  \newauthor James S. Bullock$^1$ \\
  \noindent$\!\!$ $^1$Center for Cosmology, Department of Physics and Astronomy,
  University of California, Irvine, CA 92697, USA \\
    \noindent$\!\!$ $^{2}$Department of Astronomy and Joint Space-Science Institute,
    University of Maryland, College Park, MD 20742-2421, USA }
  
\begin{document}

 \pagerange{\pageref{firstpage}--\pageref{lastpage}} 
 \pubyear{2014}

\maketitle
\label{firstpage} \begin{abstract} We study dwarf satellite galaxy quenching
  using observations from the Geha et al. (2012) NSA/SDSS catalog together with
  $\Lambda$CDM cosmological simulations to facilitate selection and interpretation.  We
  show that fewer than $30 \%$ of dwarfs ($M_\star \simeq 10^{8.5-9.5}
  \msun$) identified as satellites within massive host halos ($M_{\rm host} \simeq
  10^{12.5 - 14} \msun$) are quenched, in spite of the expectation from
  simulations that half of them should have been accreted more than 6 Gyr ago.
  We conclude that whatever the action triggering environmental quenching of
  dwarf satellites, the process must be highly inefficient.  We investigate a
  series of simple, one-parameter quenching models in order to understand what is
  required to explain the low quenched fraction and conclude that either the
  quenching timescale is very long ($> 9.5 \gyr$, a ``slow starvation'' scenario)
  or that the environmental trigger is not well matched to accretion within the
  virial volume.  We discuss these results in light of the fact that {\em most}
  of the low mass dwarf satellites in the Local Group are quenched, a seeming
  contradiction that could point to a characteristic mass scale for satellite
  quenching.\end{abstract}


\section{Introduction}
\label{sec:intro}

The cessation of star formation in galaxies is a well studied and yet poorly
understood phenomenon. That galaxies, when placed on a color-magnitude diagram,
can roughly be divided into a red, non-star-forming sequence of mostly
ellipticals and a blue, star-forming cloud of mostly spirals has been observed
at least out to $z \sim 1$ \citep{Baldry:2004cd, Bell:2004vn, Borch2006,
  Cooper2007}. It is also observed that while the number density of
star-forming galaxies is relatively constant at $z < 1$, non-star-forming (or
quenched) galaxies have roughly doubled in number density over the past
$7$--$10\,\gyr$ \citep{Bell:2004vn, Borch2006, Faber:2007ef, Bundy:2007gh,
  Brown:2007ij}.   
  
Environment appears to play a significant role in star formation quenching, and it may be the dominant driver at the lowest mass scales \citep{Cooper:2006fk, Weinmann:2006fk, Capak:2007kx, Peng:2010dq, Geha:2012kx, Woo:2013w}.  Studies comparing satellite galaxies to isolated field systems of similar stellar mass in the local Universe find that satellites tend to exhibit lower star-formation rates, more bulge-dominated morphologies, as well as older and more metal-rich stellar populations \citep{Baldry:2006b, vandenBosch:2008v, Yang:2007, Cooper2010, Pasquali:2010p}.  This observed suppression of star formation in satellite galaxies is commonly referred to as ``environmental quenching''.  

Proposed quenching mechanisms that operate preferentially on satellite galaxies
within dense environments (such as groups and clusters) range from
``strangulation" -- when the larger potential well of the host dark matter halo
accretes all of the gas that would have originally fueled star formation in the
satellite galaxy \citep{Larson:1980ve, Kawata:2008ab}, to ``harassment" -- by which close encounters between densely packed cluster or group members strip gas from around the interacting galaxies \citep{Moore:1996kl}, to ``ram-pressure stripping" -- where the cold dense gas at the center of the satellite is violently removed
from the galaxy as a result of a high-speed interaction with the hot gas halo of
the host \citep{Gunn:1972g, Bekki:2009b}. 

Determining exactly how environmental quenching proceeds is complicated by the
fact that some galaxies in the field are quenched. For isolated galaxies, free
of the influence of high-density environments (i.e.~a more massive parent halo),
the quenched fraction increases significantly with stellar mass
\citep{Peng:2010dq, Weinmann:2006fk}. This suggests that galaxies, at least
massive galaxies, quench their own star formation independent of their
environment. These observational results are supported by hydrodynamical
simulations of isolated systems, in which star formation is suppressed owing to
the formation of a merger-induced stellar spheroid that stabilizes the disk to
the collapse of molecular clouds \citep{Dekel:2009, Martig:2009ly}. This possible quenching mechanism is unique in that it depends on the mass of the stellar spheroid rather than that of the parent dark matter halo (i.e.~environment). Owing to
the fact that self- and environmental quenching primarily affect the quenched
fraction at different mass scales (with environmental quenching playing a
primary role in low stellar mass systems, and mass quenching primarily affecting
galaxies with a higher stellar mass), several authors have argued for a
separation of quenching at low redshift into two functions -- one of environment
and one of stellar mass \citep{Kovac2014, Peng:2010dq}. Disentangling these two potential quenching regimes and identifying the associated physical mechanisms that operate within them remains a challenge for modern studies of galaxy evolution at low and intermediate redshift.

In overcoming these challenges, it is therefore desirable to model {\em both} of these quenching mechanisms. Several studies have attempted to model the effects of both self- and environmental quenching using semi-analytic prescriptions. The resulting models typically associate quenching with various properties of a galaxy, such as the time since it fell into its host, in order to match observed quenched fractions as a function of stellar mass while also reproducing the observed stellar mass function \citep{Wetzel2013, Cohn2013}. However, this overlapping of effects at stellar masses greater than $\sim 10^{10}~\msun$ continues to make a separate analysis of each mechanism difficult.

Dwarf galaxies offer a particularly useful avenue for disentangling the effects
of environmental and self-quenching. As shown by \citet[hereafter G12]{Geha:2012kx}, effectively all isolated dwarf galaxies are star-forming, suggesting that self-quenching for these systems is negligible. Specifically, G12 found that non-star-forming (or passive) galaxies with stellar mass below $\sim 10^9\,\msun$ are absent beyond $\sim 1$--$1.5\,\mpc$ of a luminous neighbor in the local Universe \citep[see also][]{Wang2009}. The observations suggest that dwarf galaxies at these low masses do not shut down their own star formation; instead quenching requires the presence of a more massive neighbor. These galaxies therefore provide an ideal experimental sample: they do not self-quench, so any observed quenching can be directly related to environmental influences.

In addition to isolating the impact of environment on low-mass galaxies, the
analysis by G12 provides well-defined, quantitative observational constraints on
potential quenching mechanisms. For example, G12 probe the physical extent of
environmental quenching, finding that dwarf galaxies are only quenched within a
projected distance of $\sim 1$--$1.5\,\mpc$ of a luminous galaxy. They interpret this result by suggesting that dwarfs are being quenched at $2$--$4$ times the virial radii of their
Milky Way size hosts.  Below, we use cosmological simulations to suggest an alternative scenario in which the $\sim 1-1.5 \, \mpc$ scale more likely corresponds to the virial radius of galaxy groups or small clusters.  Moreover, G12 also find that even within $250\,\kpc$ of a luminous neighbor, the fraction of quenched systems peaks at $\sim 25$--$30\%$. The implication of these findings is that satellite quenching is remarkably inefficient at low stellar masses, and we can constrain quenching models through the requirement that they match this low efficiency.

In what follows we compare mock observations of a large $N$-body simulation to
the observed quenched fractions vs. projected distance to a luminous neighbor from G12. We consider several one-parameter models for quenching, including a simple
quenched-at-infall scenario as well as models where quenching depends on
the host $\vmax$, satellite infall time, or the ratio of the satellite's current
$\vmax$ to the $\vmax$ it had at infall. In Sections \ref{sec:obs} and
\ref{sec:sim} we describe the G12 observational sample and our simulations,
respectively.  Section \ref{sec:fquench} presents our principal results.  Finally, we summarize our findings in Section \ref{sec:summ} and discuss them in light of past work in Section \ref{sec:disc}.

\section{Observations}
\label{sec:obs}
Our observational comparisons rely on the work of G12, who construct a dwarf
galaxy sample selected from the NASA-Sloan Atlas (NSA) of the SDSS Data Release
8 spectroscopic catalog with an improved background subtraction technique
\citep{Blanton:2011b, Aihara2011}.  G12 investigate the fraction of quenched
dwarf galaxies as a function of projected distance to a more luminous
neighboring galaxy with a velocity offset of $1000\,\kms$ or less. The dwarfs in
this sample have stellar masses that range from $7.5 < \log(\mstar / \msun) <
10$. The luminous neighboring galaxies are selected from the 2MASS Extended
Source Catalog \citep{Skrutskie2006} and have stellar masses of $\mstar > 2.5
\times 10 ^{10} \msun$. We investigate a subset of this sample for simplicity,
focusing on the stellar mass bins $8.25 \leq \log(\mstar/\msun) < 8.75$ and $9.25 \leq \log(\mstar / \msun) < 9.625$.
 
G12 define quenched galaxies as having both no $\rm H\alpha$ emission and ${\rm D}_n4000 > 0.6 + 0.1\, log_{10}(\mstar/\msun)$, a criteria based on the light-weighted age of the stellar population.  We adopt the same definition of quenched for the observations in this work. See G12 for more details on the observational sample.

\section{Simulations}
\label{sec:sim}

We use the Millennium II Simulation \citep[\msii, ][]{Boylan-Kolchin:2009ly} to
construct a mock galaxy catalogue with the goal of mimicking, via abundance
matching, the sample used in G12. \msii~is a dark-matter-only simulation of
$2160^3 \approx 10$ billion particles in a box of size $L_{\rm box} =
100~\hmpc$, with a particle mass of $m_{\rm p} = 6.885 \times 10^6 \hmpc$ and a
Plummer-equivalent force softening of $\epsilon = 1~\hkpc$ in comoving
units. The cosmological parameters are $\Omega_{\rm tot} = 1.0$, $\Omega_{\rm m}
= 0.25$, $\Omega_{\rm b} = 0.045$, $\Omega_{\Lambda} = 0.75$, $h = 0.73$,
$\sigma_8 = 0.9$, and $n_{\rm s} = 1$, where $\sigma_8$ is the rms amplitude of
linear mass fluctuations in $8~\hmpc$ spheres at $z=0$ and $n_{\rm s}$ is the
spectral index of the primordial power spectrum. The simulation stores all
gravitationally self-bound dark matter subhalos down to 20 particles, which
corresponds to a resolved mass of $1.38 \times 10^8\,\msun$. Thus, \msii~is more
than adequate to make realistic comparisons to our data, as it has the
resolution required to resolve the subhalos that would likely host the galaxies in the
G12 sample.

In the figures that follow, the simulation data is constructed by selecting halos that lie in two distinct $\vmax$ (taken at infall for subhalos) bins. The lower (upper) $\vmax$ bin contains halos with $80~\kms~<~\vmax~<~90~\kms$ ($100~\kms~<~\vmax~<~110~\kms$) and roughly corresponds, via abundance matching \citep{BoylanKolchin2012}, to the G12 stellar mass bin centered at $\mstar = 10^{8.5}~\msun$ ($\mstar = 10^{9.5}~\msun$). However, rather than make an explicit comparison between a single $\vmax$ bin and a single bin in stellar mass in the plots, we attempt to encompass the uncertainty of abundance matching at low $\vmax$ by comparing points from the G12 data to shaded bands that show results (for example, the fraction of the dwarf halos that are subhalos) as we vary $\vmax$ between these two selection bins. In practice, our results are not strongly sensitive to the precise range of $\vmax$ corresponding to the satellite stellar masses from the NSA. This is evidenced by the fact that these bands are relatively thin in the following figures.

In selecting subhalos, we use the $\vmax$ at infall, hereafter $\vinfall$, where infall is defined in \msii~as the time at which a halo most recently became a subhalo (see Section \ref{sec:models}). This ensures that any tidal stripping that a subhalo has experienced will not introduce biases in the abundance matching for that halo. \citet{Behroozi:2013uq} have recently shown that $\vinfall$ is distinct from
$\rm V_{peak}$ (the largest value of $\vmax$ in the subhalo's history), that $\rm V_{peak}$ is typically set by a $1$:$5$ merger, and that the time at which it occurs does not correspond to the time at which the halo mass peaks. However, $\vinfall$ does correspond to the peak halo mass. Thus, we consider the $\vmax / \vinfall$ ratio to be a better proxy for mass loss of a subhalo than would be $\vmax / \rm V_{peak}$. We select only those halos that have a $ \vmax / \vinfall > 3/8$, ensuring that our sample is complete down to our lower $\vinfall$ limit of $80\,\kms$, as \msii~is complete down to a current $\vmax$ of $40\,\kms$.

For simplicity and without loss of generality, we mock observe the simulation by placing the ``observer" at the origin. Then, following G12, we determine the projected distance, $\rm d_{Neighbor}$, from
each halo in our dwarf mass range to the closest (in projected distance) halo that would be likely to host a more luminous galaxy. We use the term ``luminous neighbor'', once again following G12, but will use quotation marks when we are actually referring to the dark matter halo of the luminous neighbor. We choose a minimum $\vinfall$ of $150\,\kms$ for the population of ``luminous neighbors" which corresponds -- according to the abundance matching relation of \citet{BoylanKolchin2012} -- to the stellar mass limit $(> 10^{10.4} \msun)$ used to select luminous neighbors in G12. In our mock observations, there are instances where the ``luminous neighbor" of the dwarf is not the dwarf's actual host, but a subhalo of larger halo. Therefore, we use the $\vinfall$ of the ``luminous neighbors" instead of their $\vmax$ unless we are referring to a subsample of the ``luminous neighbors" for which all of the objects are actual host halos in the simulation.

In order to match the observations of G12 so that we can make appropriate comparisons, we follow exactly their method of removing contaminants in the sample. We remove halos that are close in projection but distant in velocity space by making sure that the velocity offset between the dwarf and the ``luminous neighbor" is less than $1000\,\kms$. This cutoff will indeed remove some interlopers, but as \citet{Phillips2014} show, even with a more restrictive velocity offset maximum of $500\,\kms$ and the imposition of a set of isolation criteria designed to remove groups and cluster halos, false pairs are still quite common. For LMC size satellites within $350~\kpc$ of  isolated Milky Way analogs, \citet{Phillips2014} show that the ``host-satellite" pairs are still false pairs $25\%$ of the time. This suggests that with our $1000\,\kms$ cutoff and no isolation criteria, we will quite often identify a ``luminous neighbor" that is not the actual host of the dwarf, even when the pairs are close in projection. Nonetheless, we use the $1000\,\kms$ velocity offset in order to exactly match what was done for the observations from G12 so that our comparisons to their data are meaningful.

Although \citet{Guo:2011} and \citet{Moster2013} have constructed mock
catalogues of galaxies with star formation rates and colors derived from
semi-analytic models applied to \msii, we have independently confirmed results \citep{Weinmann2006, Kimm2009, Wang2012, Wang2014} showing that the semi-analytic models over-predict the red fraction of satellites significantly. By relying instead on our simple models applied to subhalos, we aim to gain insight into the basic prescriptions that will be required to
match the data more effectively in the future.

\section{Results}
\label{sec:fquench}
\subsection{Quenched Fraction vs. Subhalo Fraction}
\label{sec:nature}

\begin{figure*}
\centering
\begin{minipage}{\columnwidth}
	\centering
	\includegraphics[scale=0.48, viewport=0 0 600 410]{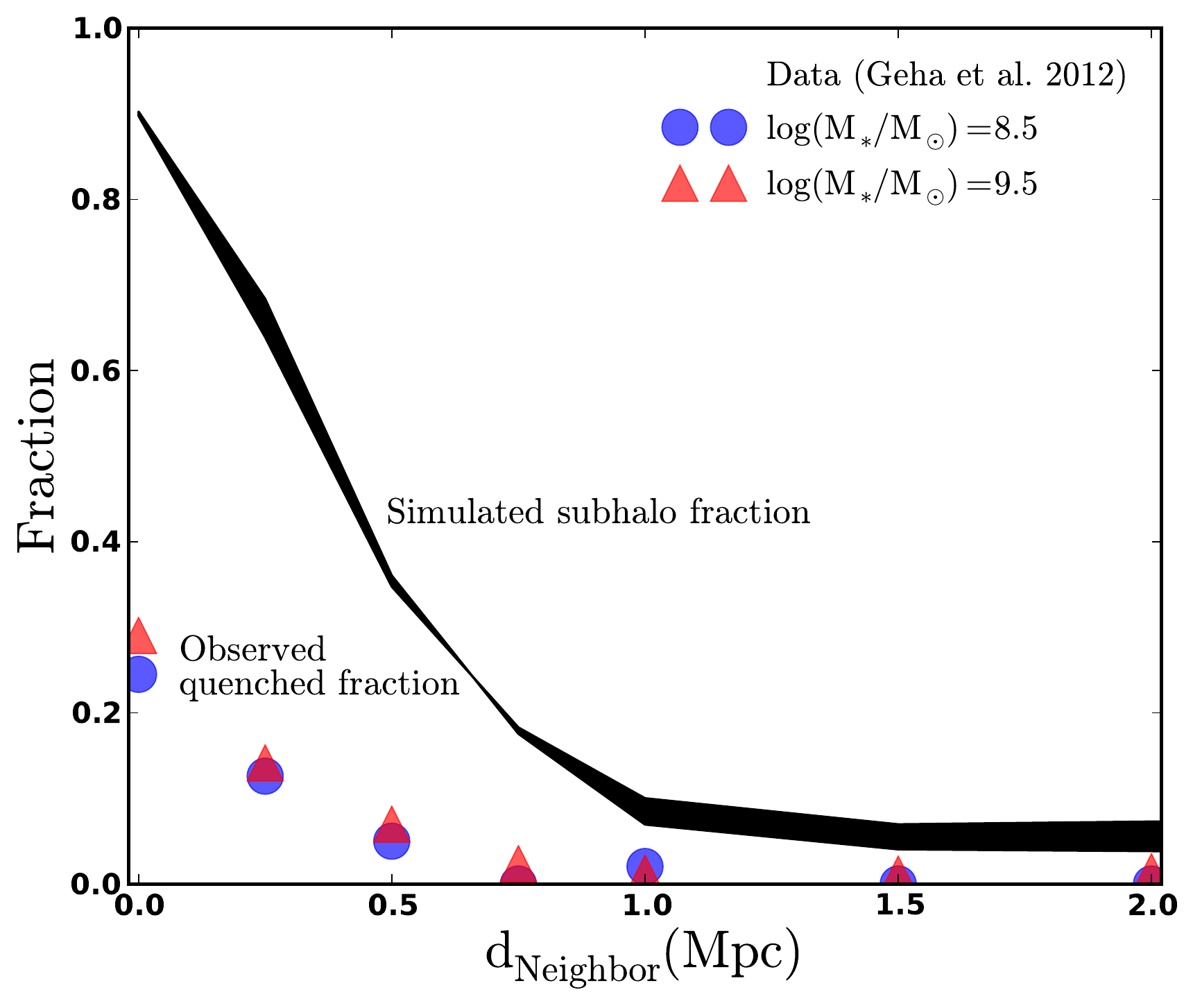}
 	\caption{The fraction of quenched dwarf galaxies as a function of the
          projected distance from the nearest luminous neighbor (symbols,
          reproduced from G12).  For comparison, the black band shows the
          fraction of those dwarf galaxies that are expected to reside within subhalos of a
          larger host, plotted as a function of the same separation measure, as
          determined by mock observations in the \msii~cosmological simulation.
          The thickness of this band corresponds to a range of $\vmax$ choices
          for identifying dwarf halos, as discussed in \S 3. The subhalo
          fraction is always well above the quenched fraction, meaning that it
          is impossible for all subhalos to be quenched; rather, satellite
          quenching at these mass scales must be fairly inefficient.  }
 \label{fig:marla}
\end{minipage} 
\hfill
\begin{minipage}{\columnwidth}
	\centering
	\includegraphics[scale=0.48, viewport=0 0 600 410]{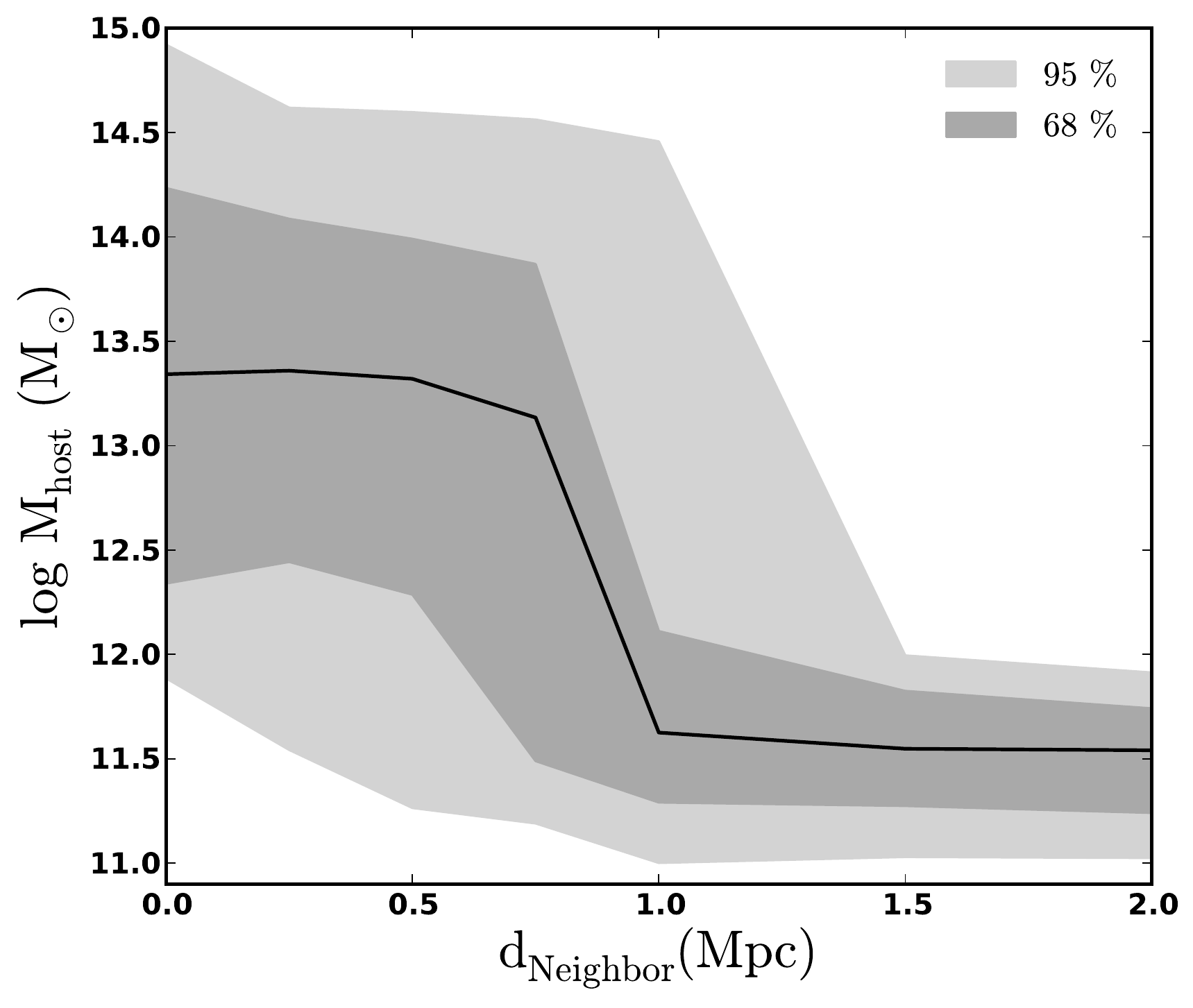}
        \caption{Range of virial masses ($95\%$, light grey; $68\%$, dark grey) for
          the actual hosts of subhalos identified in Figure 1. Hosts of dwarfs found close to
          luminous neighbors have a range of viral masses spanning the group to
          cluster scale $\sim 10^{12.5-14}\,\msun$, while hosts of dwarfs found
          far from large neighbors are concentrated around a much lower mass,
          $10^{11.5}\,\msun$. The sharp decrease in the quenched fraction below
          $\sim 0.75-1$ Mpc seen in Figure 1 is likely associated with the
          virial radius scale of groups or small clusters.}
 	\label{fig:frac_mvir}
\end{minipage}
\end{figure*}

The symbols (circles and triangles) in Figure \ref{fig:marla} reproduce results
from G12: the fraction of dwarfs that are quenched is plotted as a
function of projected distance from their nearest luminous ($\mstar > 2.5 \times
10 ^{10} \msun$) neighbor. As emphasized in G12, the quenched fraction is
effectively zero at large $\rm d_{Neighbor}$ separation, rising to $\sim 25$--$30\%$
at the smallest separations. For comparison, the black band shows the
fraction of dwarf halos in the simulation -- identified by mock observations that mirror exactly those used by G12 to produce the data points (see Section \ref{sec:sim}) -- that are known subhalos (i.e. that are subhalos within a larger FOF group).

Figure~\ref{fig:marla} reveals at least two interesting points for elaboration.
First, the subhalo fraction remains nonzero even at very large $\rm
d_{Neighbor}$ separation, hovering just under $10\%$ in a region of the figure
that is designed to target isolated galaxies. The reason for this is that
galaxies at large separation from {\em luminous} neighbors can nevertheless be
subhalos of dimmer hosts that fall below the luminosity cut (for our purposes,
halos with $\vmax < 150\,\kms$; as demonstrated explicitly in Figure 2 and  discussed
below). The fact that observed dwarfs at these large $\rm d_{Neighbor}$
separations are {\em all} star-forming, even though $\sim5$--$10\%$ of them are
identified as subhalos of {\em something}, immediately demands that not all
subhalos are quenched.  A second point of note in Figure 1 is that the subhalo
fraction rises to $\sim 90 \%$ at small $\rm d_{Neighbor}$, while the quenched
fraction remains relatively low by comparison ($\sim 25$--$30\%$).  We see again that
the mere act of being a subhalo cannot result in immediate quenching.
 
Figure \ref{fig:frac_mvir} provides a more detailed examination of the mock
observations used in Figure \ref{fig:marla}, concentrating on the subset of
halos in each $\rm d_{Neighbor}$ bin that are identified as subhalos. The
shaded bands show the range of host virial masses inhabited by subhalos for a given
$\rm d_{Neighbor}$ separation.\footnote{We emphasize that the halo mass shown is
  that of the true host halo identified in the simulations, which is not
  necessarily that associated with the ``luminous neighbor" identified in the
  mock G12 observations.}  The median halo mass is plotted as a solid black
line, while the $68\%$ and $95\%$ regions are shown in dark and light grey,
respectively.  As expected, dwarf halos that are identified as subhalos with
$\rm d_{Neighbor}$ $\gtrsim 1~{\mpc}$ have much smaller host masses than the
dwarfs with $\rm d_{Neighbor} < 0.5~\mpc$.  Dwarfs within $\sim 0.5$ Mpc are
found preferentially in cluster size host halos. Although we cannot rule out the hypothesis of G12 that the drop-off seen at $\sim 1$ Mpc corresponds to
$2$--$4$ times the virial radius of a typical $\mstar \sim 3 \times 10^{10} \msun$
galaxy, we find this scenario unlikely as it would imply that satellites are being quenched at equal or greater efficiency where the ambient gas density is very low relative to satellites within the virial radius of clusters or large groups. Thus, we find that a natural alternative explanation for the quenched fraction drop-off at $\sim 1$ Mpc is that it is set by the typical virial radius of a large group or a small cluster.

The preceding discussion has shown that satellite quenching is far too
inefficient to be caused simply by a galaxy becoming a satellite (i.e.~falling
into another dark matter halo). The fraction of observed quenched dwarfs in the
inner bin of Figure \ref{fig:marla} (within $250 \kpc$ of a more luminous
neighbor) is just over $25\%$, while over $90\%$ of simulated dwarfs selected
in the same manner and at the same projected distance are subhalos (typically, of
$\sim 10^{13.5} \msun$ hosts). This implies that at most $\sim 30 \%$ of
subhalos are quenched. This disparity between subhalo fraction and observed
quenched fraction is consistent with known problems faced by many models of galaxy formation in explaining observed satellite red fractions. In particular, models in which gas is instantaneously stripped from an infalling satellite upon entering into the host's virial radius over-predict satellite red fractions \citep[e.g.][]{Kimm2009}. In the next section, we explore a few simple models in order to understand what is required to explain the relatively low fraction of quenched dwarf satellites seen in the G12 data.

\subsection{Testing Simple Models for Quenching}
\label{sec:models}

Figure \ref{fig:frac_mvir} shows that the host halos of the star-forming dwarfs at
large $\rm d_{Neighbor}$ are systematically less massive than those of the
dwarfs found within $1\,\mpc$ of a luminous neighbor.  Since dwarfs at these
large distances are uniformly star-forming, this suggests a simple model that
limits quenching to hosts above a minimum halo mass. A model of this kind might
be motivated by the transition mass at which a quasi-static gaseous corona
forms \citep{Keres:2006, Birnboim:2007bh, More:2010kx}. However, several authors have argued that trends between the quenched fraction and host mass could be strongly affected by ``pre-processing", whereby the satellites falling into massive hosts have previously fallen into less massive host halos that are then accreted onto the more massive systems, and thus have been preferentially quenched by the first host that they fell into \citep{vandenBosch:2008v, deLucia2012,  Wetzel2013}.

\begin{figure}
 \centering
 \includegraphics[scale=0.48, viewport=0 0 800 410]{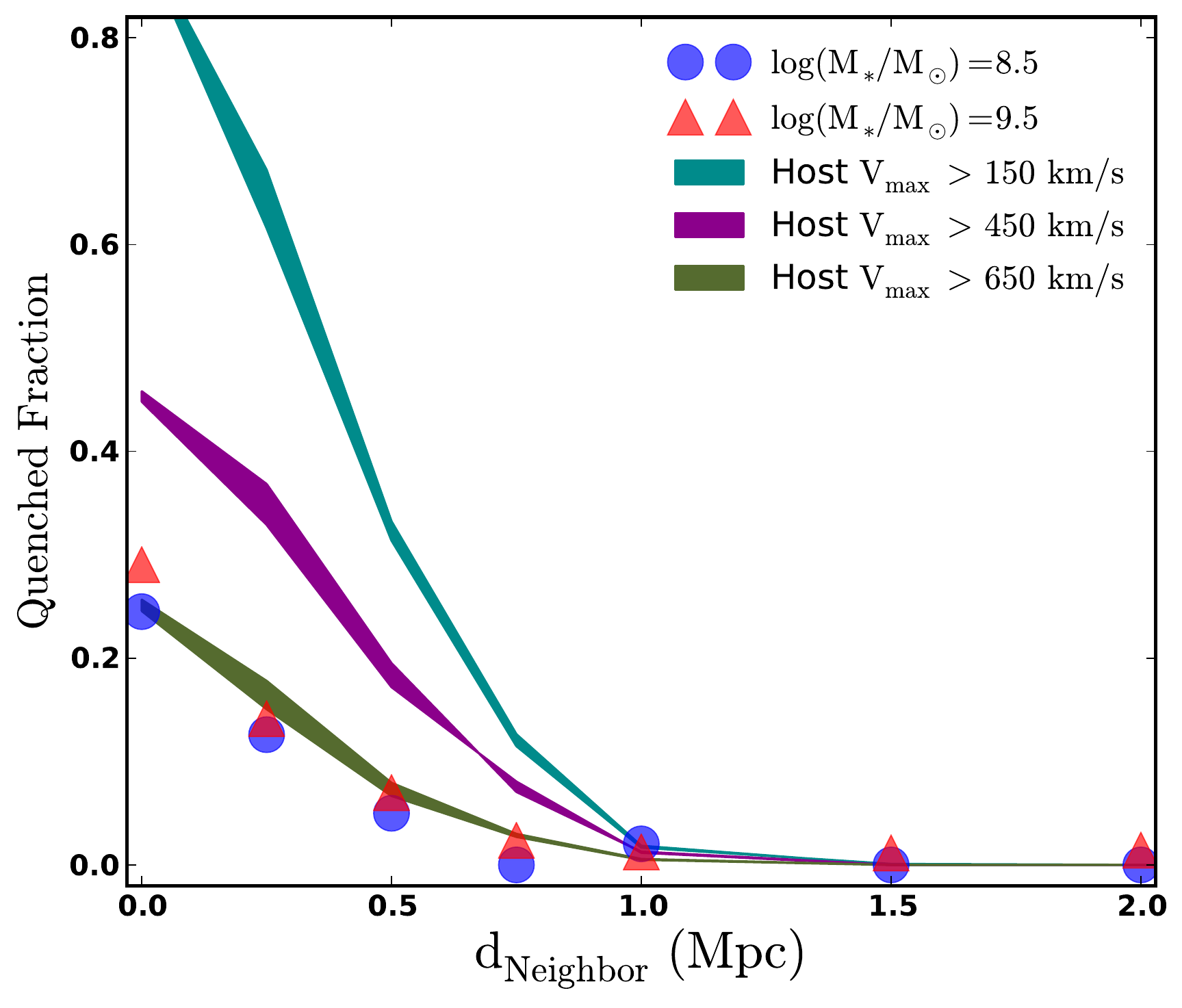}
 \caption{The relationship between quenching and host halo mass.  Colored bands show the predicted quenched fraction in a model where subhalos become quenched only when their host mass is above a given threshold: $\vmax > 150\,\kms$ (cyan), $> 450\,\kms$  (magenta), and $> 650\,\kms$ (green). 
 The symbols are the same data points shown in Figure 1.  Reproducing the data
 requires a minimum host $\vmax$ of $650\,\kms$, which is much too large
 given existing constraints on satellite quenching in lower mass
 hosts. The thickness of the bands illustrate how our results change as we
 vary the $\vinfall$ range used in identifying dwarf galaxy halos, as discussed in
 \S 3.  
 \label{fig:hvmax}
}
\end{figure}

To investigate the possibility that satellites are only quenched when falling
into a host above a certain halo mass, we develop a ``minimum mass'' quenching
model for which the resulting dependence of quenched fraction on projected
distance is shown in Figure \ref{fig:hvmax}. In this model, we define a dwarf to
be quenched once it joins the fof group of a host more massive than a
threshold $\vmax$, which we use as a proxy for host halo mass. If we set this
minimum host $\vmax$ to be the minimum $\vinfall$ required for a halo to be
considered a ``luminous neighbor", $\vinfall > 150\,\kms$ (shown as a cyan band), we can easily reconstruct the observed quenched fraction of $\rm 0$ at $\rm
d_{Neighbor} > 1-1.5~Mpc$. This is a result of the fact that subhalos this far
from their ``luminous neighbors" have hosts that are not luminous enough to fall
into the ``luminous neighbors" category. If the actual hosts were luminous enough, we would have identified them as the ``luminous neighbor", and the $\rm d_{Neighbor}$ would reflect this shorter distance to the actual host. This model does, however, have significant trouble reproducing the observed quenched fraction at small $\rm d_{Neighbor}$, with the predicted quenched fraction exceeding $90\%$ for dwarfs with $\rm d_{Neighbor} < 250~\kpc$, even when we require the host to have $\vmax > 150\,\kms$.

Only when we restrict quenching to satellites within host halos of $\vmax > 650\,\kms$ is our model able to match the observed quenched fraction. However, this is an unrealistically high value for a $\vmax$ threshold. Several studies have shown that satellites are quenched in excess of the field when seen around hosts with much smaller halo masses \citep{More:2010kx, Monachesi2011,Tollerud:2011}.  For example, using a carefully designed sample to target Milky Way size hosts, \citet{Phillips2014} find that LMC size satellites are quenched with an excess of roughly $16\%$ compared to galaxies in the field.  Furthermore, our own Milky Way, with a (dark matter) $\vmax$ less
than $220 \,\kms$ (excluding the disk contribution), has many quenched satellites,
as does M31 \citep{Mateo1998,McConnachie:2012}.  We conclude that a simple model whereby a subhalo becomes quenched only after falling within a massive ($\vmax > 650 \kms$) host is unrealistic.

It is, perhaps, more plausible that infall time is the primary factor in
determining whether a satellite becomes quenched. Models of this kind are
popular and physically motivated \citep{deLucia2012, Wetzel2013}. Because
infall time is linked to distance from the center of the potential well in the
host halo, infall time models naturally reproduce observed gradients in color
with cluster-centric distance (see \citealt{Hearin:2013}, who demonstrate this qualitative result with a more sophisticated model, and \citealt{Smith2012}).

We explore infall time ($\rm \tau_{infall}$) as a quenching parameter by associating subhalos with quenched galaxies based on the time they most recently became a subhalo of another halo in the simulation. In the \msii~database, the last time at which a halo becomes a subhalo is determined by the parameter {\em infallSnap}: the most recent {\em snapnum} at which the subhalo went from being at the center of its own friends--of--friends group to being inside another halo's friends--of--friends group. In our $\rm \tau_{infall}$ model, then, a galaxy becomes ``quenched" after falling into another galaxy's friends--of--friends (FOF) group and orbiting for a minimum amount of time. Because we use this definition of infall time, in cases where the initial crossing was followed by a subsequent pass outside the FOF group, we specifically associate infall time with the {\em last time} the subhalo joined the FOF group. However, because the definition is based on FOF group instead of virial radius, some subhalos that pass outside the virial radius of their host but stay within the FOF group will still be counted as subhalos, with an infall time dictated by when they first fell into their host. This should account for some of the ``backsplash" galaxies, which \citet{Wetzel2014} have concluded should behave very similarly to satellite galaxies in terms of quenching. In cases where the subhalo was originally accreted as a subhalo of something else (e.g. infall into a cluster as part of a group) our definition means that we track the first time the object became a subhalo to measure infall time.

\begin{figure}
 \centering
 \includegraphics[scale=0.48, viewport= 0 0 800 410]{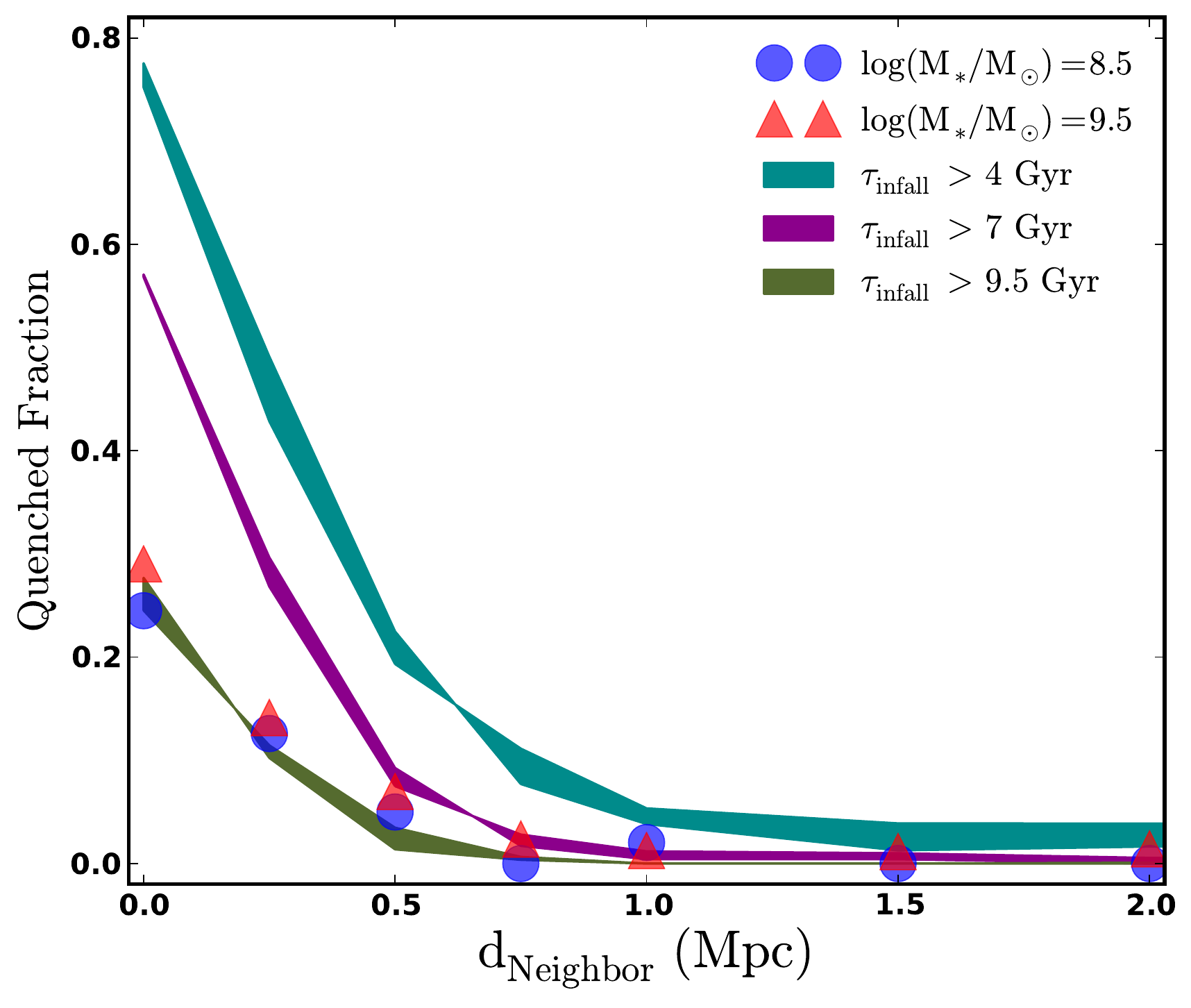}
 \caption{The relationship between quenching and infall time.  Colored bands show
 the predicted quenched fraction in a model where subhalos become quenched only
 after a time $\rm \tau_{infall}$ of orbiting within a host: $\rm \tau_{infall}$
 $> 4\,\gyr$ (cyan), $> 7\,\gyr$ (magenta), and $> 9.5\,\gyr$ (green). This
 model works only with rather long quenching timescales, $\rm \tau_{infall} 
 > 9.5\,\gyr$. The thickness of the bands correspond to different ranges of
 $\vmax$ used to identify dwarf halos, as discussed in \S 3.   
 \label{fig:tinf}
}
\end{figure}

The results of this model are illustrated in Figure \ref{fig:tinf}, where once again we are
comparing to the observed quenched fractions from G12 (blue circles and red
triangles). Here, the cyan band is the fraction of all dwarfs that are subhalos
and that became subhalos over $4~\gyr$ ago. Within
$250~\kpc$, almost $80\%$ of all dwarfs have been orbiting within their host
halos for over $4~\gyr$. Nearly $60\%$ have been orbiting for over $7~\gyr$,
shown as the yellow band. In order to reproduce the observed quenched fraction
at small separations ($\rm d_{Neighbor} < 250~\kpc$), we must restrict quenching
to only those satellites that fell into their host halo more than $9.5~\gyr$
ago. The quenched fraction that corresponds to this relatively extreme criterion
is shown as the green band in Figure \ref{fig:tinf}. This quenching timescale is very
long, and suggestive of a very inefficient quenching process, more like
strangulation than ram-pressure stripping at the virial radius. We discuss this
result in relation to other work in Section \ref{sec:disc}.

\begin{figure}
 \centering
 \includegraphics[scale=0.48, viewport=0 0 800 410]{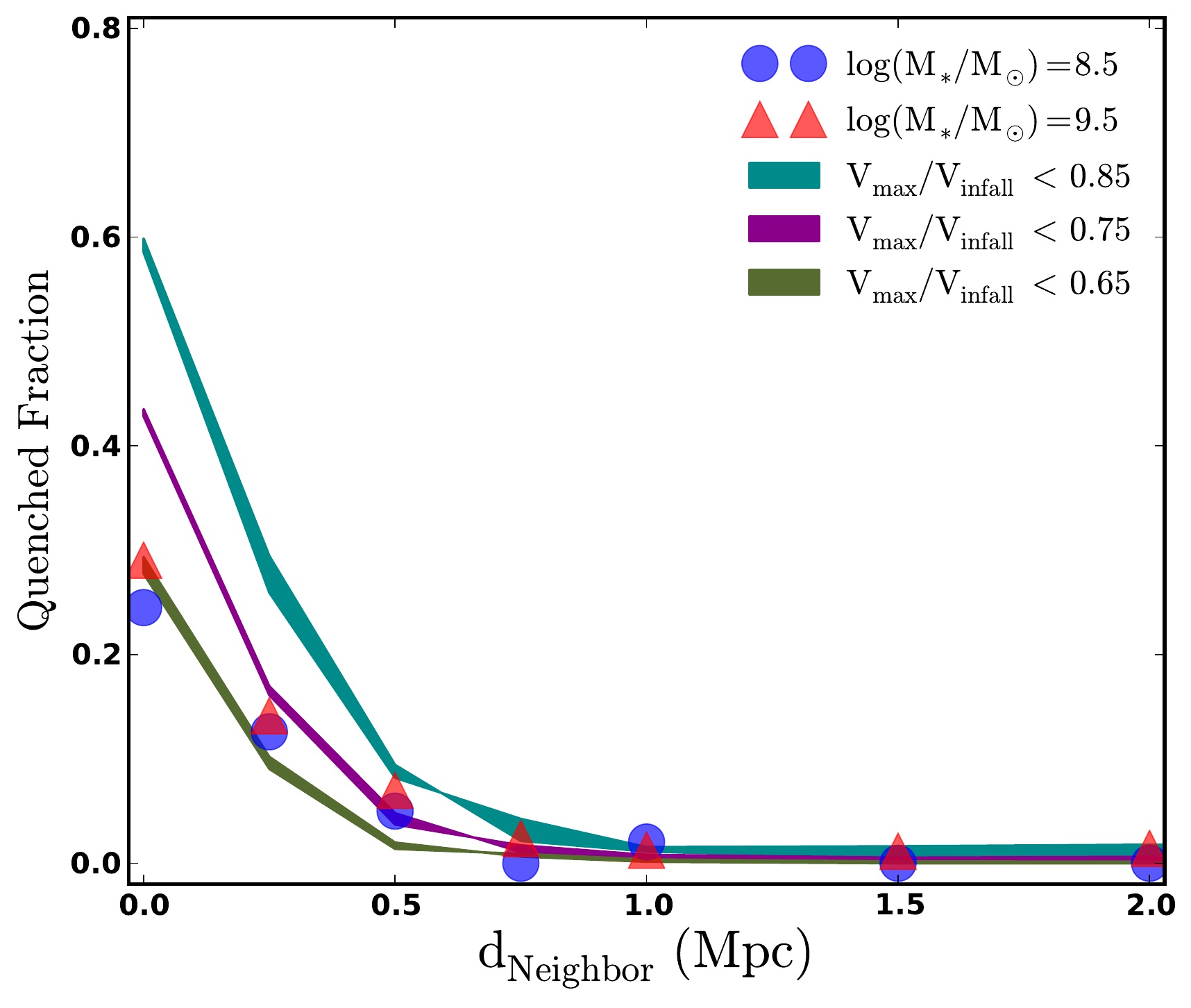}
 \caption{The relationship between quenching and mass loss.  Colored bands show
 the predicted quenched fraction in a model where satellites become quenched  when
 their subhalos have been stripped of mass beyond a certain level: $\vmax /
 \vinfall < 0.85$ (cyan), $< 0.75$ (magenta), and $< 0.65$ (green). The data are
 marginally well described by a quenching scenario with  $\vmax / 
 \vinfall = 0.65$ as the critical scale for star formation cessation.  The
 thickness of the bands correspond to different ranges of $\vmax$ used to
 identify dwarf halos, as discussed in \S 3. 
 \label{fig:vrat}
}
\end{figure}

Alternatively, subhalo quenching may have less to do with crossing the virial
radius boundary than it does with a more central encounter, where tidal forces
are greater and the hot gas density of the host halo is higher and more effective at
ram-pressure stripping.  While infall time is partially correlated with halo-centric distance, we explore a model that ties quenching to a parameter that more directly traces the tidal forces experience by a subhalo: the ratio of $\vmax$ at $z=0$ to $\vinfall$. Figure \ref{fig:vrat} shows the results of this model, with the fraction of dwarfs with current
$\vmax$ smaller than $85\%$, $75\%$ and $65\%$ of their $\vinfall$ values
plotted as cyan, magenta and green bands, respectively. Using this model to
define when a satellite galaxy is quenched, we roughly match the observed values
using a $\vmax/\vinfall = 0.65$ threshold. This corresponds to the satellite
having lost roughly $70\%$ of its infall mass, which is perhaps reasonable. At a fixed $\vmax/\vinfall$, however, this model has trouble reproducing the observed quenched fractions in detail from $0.5-1$ Mpc compared to the fixed infall time model shown in Figure \ref{fig:tinf}.

\section{Summary} 
\label{sec:summ}

We have used mock observations of the \msii~simulation
\citep{Boylan-Kolchin:2009ly} in order to interpret dwarf galaxy ($\mstar \simeq 10^9\,\msun$) quenching as a function of projected distance to a more
luminous neighbor as observed by \citet[G12]{Geha:2012kx}. Dwarfs of this mass
are particularly useful as a test bed for environmental quenching because, as
shown by G12, they are never (or almost never) quenched in isolation.

We show that while dwarfs within $250\,\kpc$ and $1000 \,\kms$ of a luminous
neighbor are subhalos of a larger host approximately $90\%$ of the time, only
about $25$--$30\%$ of such dwarfs are quenched (suggesting that $\sim 30\%$ of
subhalos are quenched at these masses). The implication is that whatever is
giving rise to subhalo quenching, the process must be fairly inefficient, at
least when evaluated relative to the subhalo population as a whole.

We investigate a model in which dwarfs become quenched only after a time
$\tau_{\rm infall}$ of being accreted into a larger host.  The required
quenching timescale is quite long, $9.5\,\gyr$ (cf. Figure \ref{fig:tinf}), compared to some estimates in the literature (see \S \ref{sec:disc}).  Alternatively, if dwarf
quenching is instead related to the tidal forces experienced by the subhalo,
then a simple model in which dwarfs with $\vmax/\vinfall < 0.65$ become quenched
does a reasonable job in reproducing the observed quenched fraction in the innermost bin of G12 (see Figure \ref{fig:vrat}), though in detail the infall time model provides a better match overall (Figure \ref{fig:tinf}). If, instead, we try to explain the relatively low quenched fraction for subhalos by demanding that only hosts larger than a critical $\vmax$ are able to quench their satellites, we find that a threshold $\vmax$ of $650\,\kms$ is required. This value is unreasonably high, as it is well above that of isolated galaxy size hosts that are known to quench at least some of their satellites \citep{More:2010kx,Monachesi2011,Tollerud:2011,Phillips2014}.

Quenching is likely a complicated process that depends on more than a
single parameter. Our aim in this work is to focus on simple models in
order to gain qualitative insight, though we also explore slightly more
complicated cases in which two conditions must be met before quenching occurs. Combining a minimum host $\vmax$ with a maximum ratio of satellite $\vmax / \vinfall$ does not significantly change the required values for either parameter. A minimum host $\vmax$ of $350\,\kms$ only moves the best fit threshold ratio to $\vmax / \vinfall = 0.70$, which is not significantly larger than the required one-parameter model value of $0.65$. We also investigate a model that requires both a minimum host $\vmax$ as well as a minimum $\rm \tau_{infall}$ for a satellite to become quenched. However, even requiring the relatively high host $\vmax$ lower limit of $350~\kms$ does not reduce the required infall time significantly. We estimate that these joint criterion for quenching will lead to a reduced threshold quenching timescale of $\sim 7.5~\gyr$.  Finally, combining $\vmax / \vinfall$ and $\rm \tau_{infall}$ does not significantly change the required values for either parameter. This is due to the large amount of overlap between subhalos that have lost a certain fraction of their $\vinfall$ and those that have been orbiting within the virial radius of their hosts for a minimum amount of time.

\section{Discussion} 
\label{sec:disc}

Applying a semi-analytic scheme that models the evolution of central and
satellite galaxies separately, while determining the quenching timescales for
each by matching the observed quenched fractions in the SDSS,
\citet{Wetzel2013} conclude that infall time is the main determinant with
regard to satellite quenching. \citet[][hereafter W13]{Wetzel2013} focus on satellite galaxies that are more massive than the dwarfs considered in this work. Their sample has a minimum satellite mass of $\mstar \sim 5 \times 10^9\,\msun$, compared to $\mstar \sim 3 \times 10^9\,\msun$ as the midpoint of the most massive stellar mass bin we consider for our dwarfs. W13 find that satellite galaxies experience ``delayed-then-rapid'' quenching, becoming quenched rapidly (within an $e$-folding time of $0.8\,\gyr$) only after having orbited their hosts for $\sim 2$--$4\,\gyr$ \citep[see also][]{McGee2009, McGee2011}. This translates into an overall quenching timescale of $\sim 3$--$5\,\gyr$, a value much less than the $9.5\,\gyr$ determined in this work.

The $3$--$5\,\gyr$ quenching timescale adopted by W13 is roughly comparable to one of our models, in which we assume a required infall time of $4\,\gyr$ before quenching; this is illustrated by the cyan band in Figure~\ref{fig:tinf}. However, this model yields a quenched fraction within $250~\kpc$ of a ``luminous neighbor" that is nearly quadruple that observed by G12. Furthermore, W13 count $\rm \tau_{infall}$ as the time
since the satellite \emph{first} fell into any host dark matter halo, while we
use the time since the satellite \emph{most recently} became a subhalo of any
host. This difference in the definition of infall time dictates that our infall
times will always be shorter than those inferred by W13 and
thereby only serves to make the contrast between their results and our work more
stark.

However, it is important to emphasize the different mass ranges investigated by W13 and our work. It is possible that invoking a simple (satellite) mass dependence in the quenching timescale could serve to largely eliminate the discrepancy between the two results. According to Figure 8 of W13, the quenching timescale becomes longer for less massive satellites. Their inferred quenching timescale nearly triples for satellites in their lower mass range, reaching
$\sim 6\,\gyr$ for satellites with $\mstar \sim 5 \times 10^9\,\msun$ compared
to $\sim 2\,\gyr$ for satellites with $\mstar \sim 10^{11}\,\msun$.

Quenching timescales found by \citet[][hereafter D12]{deLucia2012} also suggest a dependence of the quenching timescale on satellite mass. Using methods very similar to our own, they investigate mock observations of quenched fractions by testing simple models for quenching and comparing to observation. D12 find a quenching timescale of $5-7\,\gyr$ for satellite galaxies in a sample with stellar mass $10^9 < \mstar / \msun < 10^{11}$. This mass range overlaps with
ours slightly, which makes it unsurprising that they determine a timescale that
is more similar to ours and to that determined by W13 for their lowest satellite stellar mass bin. However, D12 use yet another definition for infall time: the time since a satellite first fell into its {\em current} host (we use the time it last became a subhalo of anything). By definition, the timescale employed by D12 will always be less than or equal to the timescale employed by W13, but could be shorter or longer than ours depending on the merger history of the satellite in question. For example, if a galaxy becomes a satellite of a host, and then that host iself is accreted onto a larger system, the timescale would be longer according to our definition than that of D12.

In the context of a model where infall time is the determining factor in
satellite quenching, there appears to be a qualitatively consistent trend in the
literature that lower mass satellites require longer timescales for quenching.
This result begs the question: how is it that the smallest galaxies, which are
presumably the most fragile, require the longest periods of time to become
quenched?  We speculate that this can only work in a ``slow starvation"
scenario, whereby gas-rich yet inefficiently-star-forming dwarfs continue to
form stars for a long time after their supply of fresh or recycled gas is shut
off. This possibility is consistent with the well-known fact that dwarfs have much higher gas fractions (and longer star formation timescales) than more massive galaxies \citep{Hunter1985, vanZee2001, Geha2006, Weisz2011}. If we consider the possibility that the quenching timescale is related to the gas depletion timescale, this would imply that this timescale should also increase with decreasing stellar mass. 

Observationally, however, the mass dependence of the gas depletion timescale remains fairly cloudy, especially for low-mass systems. Based on measurements of atomic hydrogen in local star-forming galaxies, gas depletion timescales are generally found to increase with decreasing stellar mass \citep{Skillman2003, Schiminovich2010}. More recent studies, however, show that star formation is a direct product of the molecular gas in a galaxy, not of all gas. In particular, on sub-kpc scales, current star formation is found to correlate strongly with molecular gas and poorly with atomic gas \citep{Wong2002, Kennicutt2007, Leroy2008, Bigiel2008}. Moreover, recent measurements of CO emission in nearby galaxies find that the molecular gas
depletion timescales are constant, or possibly even decreasing in lower stellar mass systems \citep{Leroy2008, Genzel2010, Bigiel2011, Saintonge2011, Boselli2014}. For systems with stellar mass less than $10^9~\msun$, however, the constraints are generally weak due to the difficulty of detecting CO emission in low-mass systems --- a limitation that will hopefully soon be overcome for larger samples using more sensitive facilities such as ALMA.

\begin{figure}
 \centering
 \includegraphics[scale=0.48, viewport=0 0 800 410]{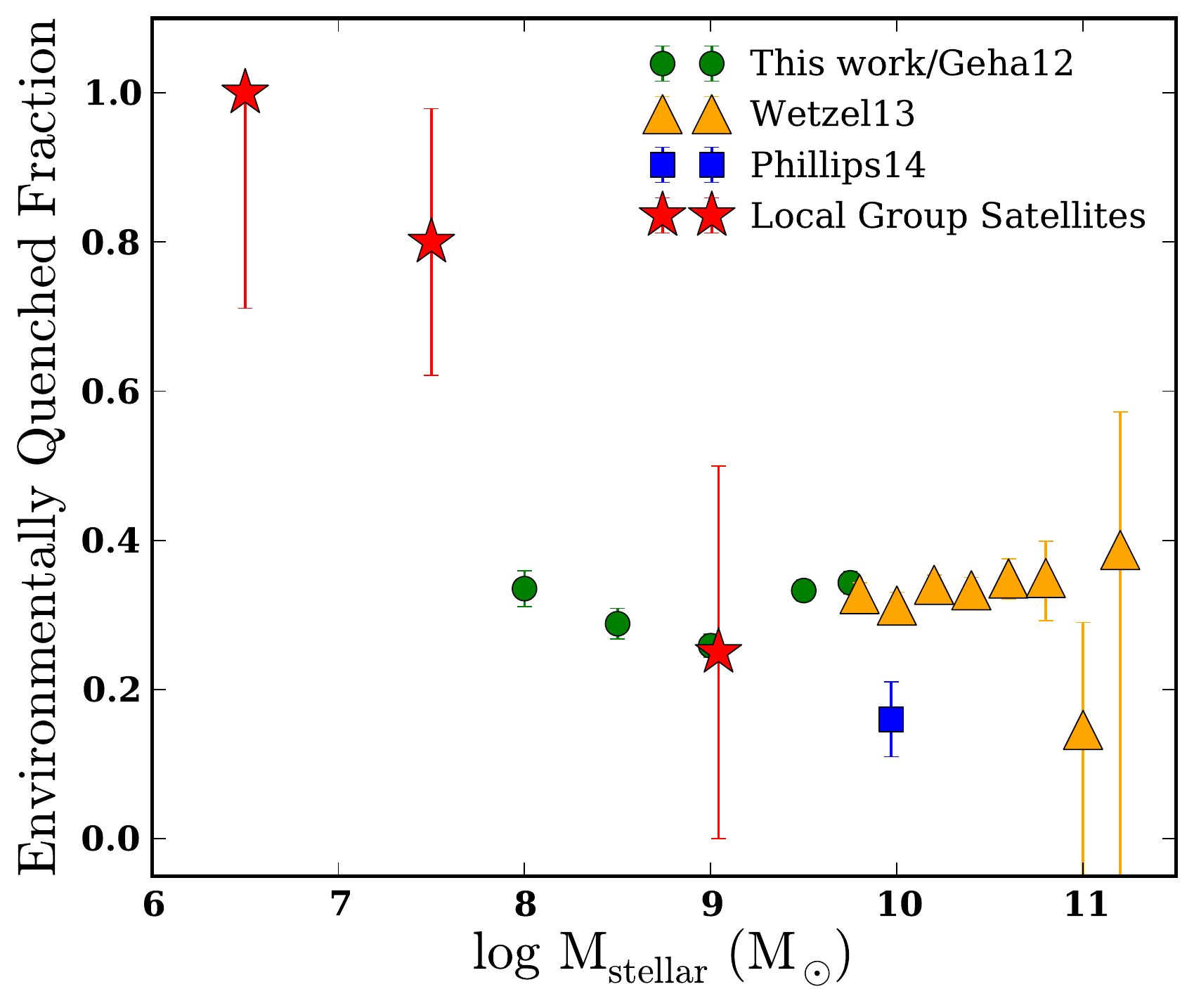}
 \caption{The environmentally quenched fraction -- the fraction of satellites that are quenched in excess of that expected in the field, i.e. the fraction of satellites that are quenched because they are satellites. We see that while environmental quenching seems to have an approximately constant efficiency of $\sim~30\%$ at stellar masses from $10^8$ to $10^{11}~\msun$, there appears to be a dramatic upturn in quenching at lower stellar masses (if the Local Group is typical).
 \label{fig:fconvert}
}
\end{figure}

Regardless of the underlying cause, our results indicate that dwarf satellites
in the $\mstar \sim 10^{8.5-9.5} \msun$ mass range are quenched only $\sim 25$--$30\%$ of the time. However, for more massive satellites there is quenching in the field, and so the quenched fraction is not the same as the fraction of satellites that are quenched because they are satellites. Thus, a proper comparison of environmental quenching over different stellar mass regimes is best made by comparing only the ``environmentally quenched fraction". This is the fraction of satellites that are quenched but would have otherwise not be quenched in the field, and so is equivalent to the overall quenched fraction for low mass galaxies. Our ``environmentally quenched fraction" is largely equivalent to the ``transition fraction", $f_{tr|s,bc}$, first introduced by \citet{vandenBosch:2008v}, as well as to the ``conversion fraction", $\rm f_{convert}$, of \citet{Phillips2014}  and the ``excess red fraction", $\rm f^Q_{excess}$, of \citet{Wetzel2013}, the latter two of which are plotted in Figure \ref{fig:fconvert} alongside our quenched fraction. As Figure \ref{fig:fconvert} shows, over the stellar mass range $10^{8}$--$10^{11}~\msun$, the environmentally quenched fraction is almost completely independent of stellar mass. The yellow triangles show results from \citet{Wetzel2013}, for which satellite galaxies reside typically in clusters. The green points are taken from our G12 sample, and include only systems with $\rm d_{Neighbor} < 250~\kpc$, which again typically reside in small clusters (Figure \ref{fig:hvmax}; and we have taken into account the fact that only $\sim~90\%$ of the galaxies in this bin are true subhalos, Figure \ref{fig:marla}).  The \citet{Phillips2014} point (blue square) is somewhat different in that these galaxies were chosen to reside within Milky-Way size hosts rather than clusters. This may explain the slightly lower environmentally quenched fraction. The results displayed here are consistent with other results suggesting that quenching efficiency is independent of stellar mass \citep{vandenBosch:2008v, Peng:2010dq, Tinker2013}.

Although the statistics for satellites in the stellar mass range $\mstar \sim 10^{8.5-9.5} \msun$ within the Local Group are very low, it is interesting that the quenched fraction for these dwarfs is not too far from $\sim 25$--$30\%$. Within this mass range, the LMC, the SMC and M33 are star-forming while NGC 205 and M32 are quenched. Once we adjust the quenched fraction to the environmentally quenched fraction by accounting for the fraction of quenched galaxies at this mass in the field, the fraction of high mass satellites of the Milky Way and M31 that have been quenched as satellites is broadly consistent with the results of this work. This can be seen in Figure \ref{fig:fconvert}, where the high mass Local Group satellites have been represented by a point placed at the mean of their stellar mass values.

This consistency of the environmentally quenched fraction over so many orders of magnitude is particularly puzzling in light of the known (very high) quenched fraction of dwarf satellites in the Local Group in the mass range just below $\mstar \simeq 10^8 \msun$ \citep{Mateo1998,McConnachie:2012}. The low mass Local Group points in Figure \ref{fig:fconvert} show a marked increase in the environmentally quenched fraction just below $10^8~\msun$. Furthermore, observations of the nearby group M81 show that almost all of the low mass satellites of that group are also quenched \citep{Kaisin2013, Karachentsev2014}. Recall that at these low stellar masses, the overall quenched fraction is equivalent to the environmentally quenched fraction, since nearly all isolated systems are star-forming. Of course, most of the quenched dwarf satellites in the Local Volume are of significantly lower stellar mass than those in G12 sample, but if stellar mass is the determining factor, it requires a reversal of sorts: rather than continuing the trend of longer quenching timescales for lower mass galaxies as discussed above, the high quenched fraction of low mass dwarfs in the Local Group suggests a sudden uptick in quenching efficiency below $\mstar \sim 10^8 \msun$ (see Phillips et al. 2014b {\em in prep.} for more on this). It is possible that we are seeing a second physical process for quenching emerging in the low-mass dwarf regime: ram-pressure stripping, which should act more efficiently on systems with shallow potential wells, might well be at work. Future observations that probe this lower stellar mass regime with greater statistical samples will be required to determine whether the physics of satellite quenching transitions at the dwarf spheroidal mass scale from processes that act inefficiently to those that squelch star formation almost uniformly.

\section*{Acknowledgments} 
We thank Marla Geha, Andrew Wetzel and their collaborators for kindly suppling their observational results in tabular form, as well as for productive discussions. CW in particular thanks Shea Garrison-Kimmel for technical support. CW and JSB were supported by NSF grants AST-1009973 and AST-1009999. We also thank Frank van den Bosch for his suggested edits as the referee, which made the paper much clearer.

\label{lastpage}
\end{document}